\documentclass[12pt]{spieman}  
\usepackage[]{graphicx}
\usepackage{setspace}
\usepackage{tocloft}
\usepackage{mathtools}
\usepackage{mathenv}

\title{Active compensation of aperture discontinuities for WFIRST-AFTA: analytical and numerical comparison of propagation methods and preliminary results with an WFIRST-AFTA like pupil.} 

\author{Johan Mazoyer \supscr{a}{}, Laurent Pueyo \supscr{a}{}, Colin Norman \supscr{b}{}, Mamadou N'Diaye \supscr{a}{}, Roeland P. van der Marel \supscr{a}{}, R\'e{}mi Soummer \supscr{a}}

\affiliation{\supscrsm{a}Space telescope Science Institute, 3700 San Martin Drive, Baltimore, MD 21218\\
\supscrsm{b}Johns Hopkins University, Zanvyl Krieger School of Arts \& Sciences, Department of Physics \& Astronomy, Bloomberg Center for Physics and Astronomy, 3400 N. Charles Street, Baltimore, MD 21218}

\cftpagenumbersoff{figure}
\cftpagenumbersoff{table} 
\begin{document} 
\maketitle 
 



\begin{abstract}
The new frontier in the quest for the highest contrast levels in the focal plane of a coronagraph is now the correction of the large diffractive artifacts effects introduced at the science camera by apertures of increasing complexity. 

Indeed, the future generation of space and ground based coronagraphic instruments will be mounted on on-axis and/or segmented telescopes: the design of coronagraphic instruments for such observatories is currently a domain undergoing rapid progress. One approach consists of using two sequential Deformable Mirrors to correct for aberrations introduced by secondary mirror structures and segmentation of the primary mirror. The coronagraph for the WFIRST-AFTA mission will be the first of such instruments in space with a two Deformable Mirrors wavefront control system. 

Regardless of the control algorithm for these multi Deformable Mirrors, they will have to rely on quick and accurate simulation of the propagation effects introduced by the out-of-pupil surface. In the first part of this paper, we present the analytical description of the different approximations to simulate these propagation effects. In Annex A, we prove analytically that, in the special case of surfaces inducing a converging beam, the Fresnel method yields high fidelity for simulations of these effects. We provide numerical simulations showing this effect.

In the second part, we use these tools in the framework of the Active Compensation of Aperture Discontinuities (ACAD) technique applied to pupil geometries similar to WFIRST-AFTA. We present these simulations in the context of the optical layout of the High-contrast imager for Complex Aperture Telescopes, which will test ACAD on a optical bench. The results of this analysis show that using the ACAD method, an apodized pupil lyot coronagraph and the performance of our current deformable mirrors, we are able to obtain, in numerically simulations, a dark hole with an WFIRST-AFTA-like pupil. Our numerical simulation shows that we can obtain contrast better than $2.10^{-9}$ in monochromatic light and better than $3.10^{-8}$ with 10\% bandwidth between 5 and 14 $\lambda/D$.
\end{abstract}


{\noindent \footnotesize{\bf Address all correspondence to}:Johan Mazoyer, Space Telescope Science Institute 3700 San Martin Drive, Baltimore, MD 21218; Tel: 410-338-4791; E-mail:  \linkable{jmazoyer@stsci.edu} }

\begin{spacing}{2}   
\end{spacing}
\section{Introduction}
\label{sect:intro}  

The current generation of high contrast coronagraphic instruments on space based or ground based telescopes were designed to reach contrast levels of $10^{-6}$. In this case, the impact of complex apertures (secondary support structures) in the performance in contrast in the focal plane of the coronagraph is negligible. Coronagraphs have thus been optimized on circularly axi-symmetric apertures, which only take into account the central obscuration (e.g. Soummer et al. 2011\cite{Soummer11}{}) and active optics systems are mostly correcting for phase aberrations introduced by the atmosphere with one high order deformable mirror (DM) in the pupil plane. 

However, the next generation of high contrast instruments will have to achieve contrast levels at least 100 times better on the ground\cite{Guyon12} and 10'000 times better in space, with pupils of increasing complexity. Indeed, the WFIRST-AFTA (Wide-Field Infrared Survey Telescope - Astrophysics Focused Telescope Assets) aperture has 6 secondary support structures and most of the future ground and spaced based primary mirrors will be segmented. High contrast imaging is now at a point where the next hurdle is to deal with diffraction from complex apertures. Another problem is the correction of amplitude aberrations (local difference of transmission by the telescope optics), which will eventually limit the contrast performance. Indeed, the simultaneous correction of phase and amplitude aberrations with a single DM can only be achieved on half of the image plane at most (Bord\'e \&Traub 2006\cite{Borde06}{}).

An elegant method to introduce apodization without limiting the throughput of the instrument is to use a set of two mirrors where at least one of them is located outside of the pupil plane. These mirrors can either be fixed, like in the family of coronagraphs derived from the phase induced amplitude apodization coronagraph (PIAA, Guyon et al. 2005\cite{Guyon05}{}) or deformable. Several 2 DM techniques have therefore been developed either to correct simultaneously for phase and amplitude aberrations (Pueyo et al. 2009a\cite{Pueyo09}{}) or to correct for the effect of complex apertures in the focal plane (e.g. active compensation of aperture discontinuities or ACAD, Pueyo \& Norman 2013 \cite{Pueyo_Normann13}{}). 

The development of these methods (in particular, determining the correct shapes of the two mirrors) required the use of numerous propagation simulations. The general equation of the electrical field at a given distance from an aperture (here, the first mirror in pupil plane) can be accurately approximated by the Huygens integral (see e.g. §8.2 in Born and Wolf 1999 \cite{born1999principles}{}). However, in the case of a two-dimensional (2D) simulation with a reasonable number of pixels (\textit{i.e.} several thousands), the numerical simulation of propagation using the Huygens integral, although very accurate, is time consuming. Several approximations have been introduced to simplify the measurement of this integral, depending on the case. 

In the favorable case of an axisymmetric aperture, the S-Huygens approximation was developed by Vanderbei \& Traub 2005\cite{Vanderbei_Traub05}{} and Belikov et al. 2006\cite{Belikov06}{}. This method is using a change into polar coordinates in the Huygens integral to obtain an accurate approximation, very fast to compute. However, in the more general case of non-axisymmetric apertures, other algorithms had to be developed. The purpose of the first part of this paper is to compare two of those approximations, the Stretched-Remapped Fresnel approximation (hereafter SR-Fresnel, \cite{Pueyo11}) and the classical Fresnel approximation (see for example Goodman, 1996\cite{goodmanfourier96}{}). This study will give a framework for the simulation of propagation in several cases, including the case of the ACAD technique. 

The purpose of the second part of this article is to demonstrate with correctly simulated 2 DM propagation that the ACAD technique can mitigate the impact of the pupil asymmetries on contrast performance and thus enable high contrast on the unfriendly pupil of WFIRST-AFTA. These numerical simulations were also realized to test the feasibility of this correction on the High-contrast imager for Complex Aperture Telescopes (HiCAT) test bench. Indeed, we demonstrate that the required shapes of the DMs that will be associated to this test will not exceed the capabilities of the test bench DMs.

\section{Analytical expressions of propagated fields} 
\label{sect:Analytical}

In this section, we compare analytically the degree of approximation of SR-Fresnel and Fresnel in regards to the Huygens integral. In Section~\ref{sect:DHI}, we present the notation and equations derived in previous articles. Sections~\ref{sect:Huygens}, \ref{sect:SRFResnel} and \ref{sect:FResnel} presents the Huygens propagation and the two approximations that we studied in this analysis. The analytical comparison of the precision of these two approximations is in Annex~A.

\subsection{Notations and previous equations} 
\label{sect:DHI}
\begin{figure}
   \begin{center}
   \includegraphics[height=7cm]{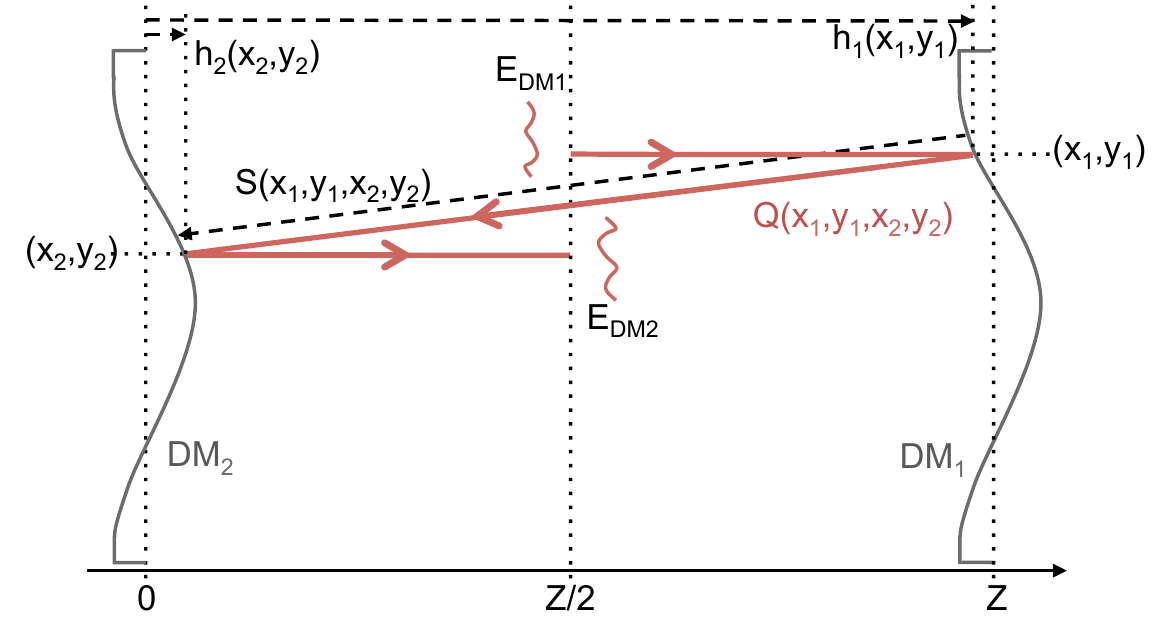} 
   \end{center}
   \caption[Notations used in this article for the 2 DM propagation] 
   { \label{fig:shema_propa}
Notations used in this article for the 2 DMs propagation, using a representation where the DMs are tip-tilt-free (e.g. using an on-axis design for the two DMs). The red solid line with arrows represents the optic ray going from $(x_1,y_1)$ to $(x_2,y_2)$, of total optical path length $Q(x_1,y_1,x_2,y_2)$.} 
   \end{figure}

In this section, we introduce our notations and present the different approximations. A beam is encountering two sequential deformable mirrors (DMs) of diameter $D$ and separated by a distance $Z$. We place ourselves in the case where the input and output beam are collimated, as in Vanderbei \& Traub (2005)\cite{Vanderbei_Traub05}{}. Figure \ref{fig:shema_propa} shows the notations used in Pueyo et al. (2011)\cite{Pueyo11}{} and in this article. We used a representation where both DMs are tip-tilt-free. We also define $h_1$ and $h_2$ so as to consider a no piston case (\textit{i.e.} in the case of two flat mirrors $h_1 = Z$ and $h_2 = 0$).

We denote $E_{DM_1}$ and $E_{DM_2}$ the electrical fields before and after encounter with the DMs. 

$Q(x,y,x_2,y_2)$ is the total optical path length between \textit{any given points} $(x,y)$ of the $DM_1$ and $(x_2,y_2)$. Using Figure, \ref{fig:shema_propa} we can write:
	\begin{eqnarray}
		\label{eq:Q}
            Q(x,y,x_2,y_2) &=& (h_1(x,y) - \dfrac{Z}{2}) + S(x,y,x_2,y_2) + (\dfrac{Z}{2} - h_2(x_2,y_2))  \nonumber \\
            &=& h_1(x,y) - h_2(x_2,y_2) + S(x,y,x_2,y_2)
    \end{eqnarray}
where $S(x,y,x_2,y_2)$ is the distance between the points on the surface of DM1 $(x,y,h_1(x,y))$ and DM2 $(x_2,y_2,h_2(x_2,y_2))$:
    \begin{equation}
	\label{eq:S}
S(x,y,x_2,y_2) = \sqrt{(x_2-x)^2 + (y_2-y)^2 + (h_1(x_2,y_2) - h_2(x,y))^2 }
	\end{equation}
We will make several approximations of the term $Q$, in order to simplify the calculation of this integral, in a fashion that allows us to use a CPU friendly Fourier Transform (hereafter FT).

At the $(x_2,y_2, h_2(x_2,y_2))$ point on the surface of the second DM, there is an unique incident ray incoming according to ray optics and considering that all rays are initially collimated. We call $(x_1,y_1)$ this position on the first DM, where this "ray optics" ray impacted the first DM. We can retrieve this location of the incident ray using the functions:
    \begin{equation}
	\label{eq:f1g1}
        x_1(x_2,y_2) = f_1(x_2,y_2)~~~~;~~~~y_1(x_2,y_2) = g_1(x_2,y_2)
	\end{equation}

Similarly, for a given point $(x_1,y_1, h_1(x_1,y_1))$ on DM1, we also introduce the functions that allow the determination of the corresponding unique associated position $(x_2,y_2)$ on DM2 according to ray optics.
\begin{equation}
    \label{eq:f2g2}
        x_2(x_1,y_1) = f_2(x_1,y_1)~~~~;~~~~y_2(x_1,y_1) = g_2(x_1,y_1)
\end{equation}

It has been demonstrated (See Vanderbei \& Traub, 2005\cite{Vanderbei_Traub05}{}) that the optical path is constant: 
\begin{eqnarray}
	\label{eq:const_quantity}
   Q_0 &=&  h_1(x_1,y_1) - h_2(f_2(x_1,y_1),g_2(x_1,y_1)) + S(x_1,y_1,f_2(x_1,y_1),g_2(x_1,y_1)) \nonumber \\
   &=& h_1(f_1(x_2,y_2),g_1(x_2,y_2)) - h_2(x_2,y_2) + S(f_1(x_2,y_2),g_1(x_2,y_2),x_2,y_2) \nonumber \\
   &=& 2Z
\end{eqnarray}
We evaluate this constant, which is equal to $2Z$ in the case of two flat DMs: $h_1 = Z$, $h_2 = 0$, $S = Z$. This property allows the derivation of the link between the shape of the DMs $\dfrac{\partial h_i}{\partial x}$ and $\dfrac{\partial h_i}{\partial y}$  and the functions $f_i$ and $g_i$:
\begin{equation}
	\label{eq:df1dxi}
       \dfrac{\partial h_1}{\partial x}\bigg|_{(x_1,y_1)}  = \dfrac{f_2(x_1,y_1)-x_1}{2Z}~~~~;~~~~\dfrac{\partial h_1}{\partial y}\bigg|_{(x_1,y_1)} = \dfrac{g_2(x_1,y_1) - y_1}{2Z}
\end{equation}
\begin{equation}
	\label{eq:df2dxi}
       \dfrac{\partial h_2}{\partial x}\bigg|_{(x_2,y_2)} = \dfrac{x_2 - f_1(x_2,y_2)}{2Z}~~~~;~~~~\dfrac{\partial h_2}{\partial y}\bigg|_{(x_2,y_2)} = \dfrac{y_2 - g_1(x_2,y_2)}{2Z}
	\end{equation}
We define the apodization function $A$ of the second deformable mirror as in Vanderbei et al. (2005) \cite{Vanderbei_Traub05}{}: 
\begin{equation}
	\label{eq:apod}
       E_{DM2}(x_2,y_2) = A(x_2,y_2).E_{DM1}(x_1,y_1)\,\,,
	\end{equation}
where $A$ is real. The conservation of energy in the different planes then imposes\cite{Pueyo11}{}: 
\begin{equation}
	\label{eq:energy_cons}
	\dfrac{\partial f_2}{\partial x}\dfrac{\partial g_2}{\partial y} - \dfrac{\partial f_2}{\partial y}\dfrac{\partial g_2}{\partial x} = A(x,y)^2
\end{equation}
Using Equations \ref{eq:energy_cons} and \ref{eq:df1dxi}, we can deduce the second order Monge-Ampere equation connecting the shape of the surfaces to this apodization:
\begin{equation}
	\label{eq:Monge_ampereh2}
	\left(1+2Z\dfrac{\partial^2 h_2}{\partial x^2}\right) \left(1+2Z\dfrac{\partial^2 h_2}{\partial y^2}\right) - \left(2Z\dfrac{\partial^2 h_2}{\partial x\partial y}\right)^2= A(x,y)^2
\end{equation}
Similarly, we have for $h_1$ :
\begin{equation}
	\label{eq:Monge_ampereh1}
	\left(1+2Z\dfrac{\partial^2 h_1}{\partial x^2}\right) \left(1+2Z\dfrac{\partial^2 h_1}{\partial y^2}\right) - \left(2Z\dfrac{\partial^2 h_1}{\partial x\partial y}\right)^2= \dfrac{1}{A(x,y)^2}
\end{equation}

\subsection{Propagation simulation and Direct Huygens integral} 
\label{sect:Huygens}

Our objective is to express the output electric field $E_{DM_2}$ at any point $(x_2,y_2)$ of the $DM_2$, using the input electric field $E_{DM_1}$ and the shapes $h_1$ and $h_2$ of the surface of the two DMs. The output electric field $E_{DM_2}$ can first be written using the Huygens integral, and using the scalar approximation :
	\begin{equation}
	\label{eq:Huygens_integral}
E_{DM_2} (x_2,y_2) = \dfrac{1}{i\lambda Z} \int_\mathcal{A} E_{DM_1} (x,y) e^{2i\pi Q(x,y,x_2,y_2)/ \lambda}dxdy
	\end{equation}
where $\mathcal{A}$ is the complete aperture. However, this integral is very time-consuming to simulate numerically in the general 2D case. Several approximations of the quantity $Q(x,y,x_2,y_2)$ can be used to simplify this integral and reduce the time of numerical simulation for a two mirror propagation. We used two main approximations:
\begin{itemize}
\item[a.] $Z \gg D$ which expresses that the two DMs are small compared to the distance between them.
\item[b.] $h_1-h_2 \sim Z$ which expresses that the mirror deformations are small compared to the distance between the DMs.
\end{itemize}
Because, when seeking to generate amplitude modulation with the two DMs, the mirror surfaces scale as $Z/D^2$, as discussed in Pueyo \& Norman (2013)\cite{Pueyo_Normann13}{} and shown above, the deviation from the ideal case of any approximation of $\dfrac{Q(x,y,x_2,y_2)}{Z}$ can be quantified as a power of $\dfrac{D}{Z}$. Next we revisit two numerical methods already presented in the literature in this framework and take a particular care to express the order of approximation in terms of power of $\dfrac{D}{Z}$.

\subsection{SR-Fresnel approximation} 
\label{sect:SRFResnel}

The SR-Fresnel, presented in Pueyo et al. (2011)\cite{Pueyo11} and developed in Pueyo \& Norman (2013)\cite{Pueyo_Normann13}{} is aiming at realizing a second order development of Q around the geometrical optic path $Q_0 = 2Z$:
\begin{eqnarray}
	\label{eq:Qsecondorder}
	\dfrac{Q(x,y,x_2,y_2)}{Z} &=& 2 + \dfrac{1}{2Z}\dfrac{\partial^2 Q}{\partial x^2}\bigg|_{(x_1,y_1)} (x-x_1)^2 + \dfrac{1}{2Z}\dfrac{\partial^2 Q}{\partial y^2}\bigg|_{(x_1,y_1)} (y-y_1)^2 \nonumber\\
	&& +\dfrac{1}{Z}\dfrac{\partial^2 Q}{\partial x\partial y}\bigg|_{(x_1,y_1)}(x-x_1)(y-y_1) + \mathcal{O}(\left(\dfrac{D}{Z}\right)^3)
\end{eqnarray}

This approximation was developed into two different methods: the classical SR-Fresnel (Equation 13 in Pueyo \& Norman, 2013\cite{Pueyo_Normann13}{}) and the Fourier SR-Fresnel (Equation 16 in Pueyo \& Norman, 2013\cite{Pueyo_Normann13}{}). It was also the approximation underlying the PASP (PIAA Angular Spectrum Propagator Pueyo et al. 2009b\cite{Pueyo_SPIE09}{}, Krist et al. 2010\cite{Krist10}{}, 2013\cite{Krist13}{}) algorithm developed to propagate efficiently arbitrary wavefront through a PIAA coronagraph. As indicated above it is an $\mathcal{O}[ (\frac{D}{Z})^3]$  approximation.

\subsection{Fresnel approximation} 
\label{sect:FResnel}
Previous authors have often presented the Fresnel solution as 
\begin{equation}
	\label{eq:Qfresnel2}
	\dfrac{Q(z,y,x_2,y_2)}{Z} \simeq 2\dfrac{h_1 - h_2}{Z} + \dfrac{1}{2} \left(\left(\dfrac{x-x_2}{Z}\right)^2 + \left( \dfrac{y-y_2}{Z}\right)^2 \right).
\end{equation}
without specifying the order of approximation of this equation. Goodman (1996)\cite{goodmanfourier96}{} presents a qualitative argument regarding the numerical validity of this approximation but did not quantify its precision in the framework of phase induced apodization. In this article, we prove that in some cases, this solution is a $\mathcal{O}(\left(\dfrac{D}{Z}\right)^4)$ approximation, and therefore is more accurate than the ones derived using Equation \ref{eq:Qsecondorder} in these configurations.

\section{SR-Fresnel or Fresnel ?} 
\label{sect:chosen_algorithm}

In Annex A, we analytically prove that when both $|max(1/A^2 -1)| \ll Z/D$ and $|max(A^2 - 1)| \ll Z/D$, the Fresnel approximation (Equation~\ref{eq:Qfresnel2}) is actually a $\mathcal{O}(\left(\dfrac{D}{Z}\right)^4)$ approximation and is more accurate than the SR-Fresnel one.

In the case of ACAD, we have $Z/D > 10$ (see next sections). The DM surfaces are adjusted so that the beam is locally converging around the secondary support structures, e.g. $2>A>1$, and very weakly diverging anywhere else ($A \sim 0.9$). Therefore, in this case, $|max(1/A^2 -1)|< |max(A^2 - 1)| \ll Z/D$. In this context, we showed in this the Fresnel approximation is more appropriate to capture the physics of diffraction between the DMs. In the next section, we provide a numerical example of this case. \\

On the other hand in the case of a PIAA coronagraph the beam is very diverging near the edge of the pupil ($A \sim 10^{-5}$ without a post-apodizer, $A \sim 10^{-2}$ with one). In this case, with $Z/D \sim 10$, the SR-Fresnel approximation is more accurate. However because it is only a $\mathcal{O}(\left(\dfrac{D}{Z}\right)^3)$ approximation, it is used in practice in conjunction with the S-Huygens approximation (a true  $\mathcal{O}(\left(\dfrac{D}{Z}\right)^4)$ to simulate a coronagraph. Indeed, in Krist et al. (2013)\cite{Krist13}{}, the bulk of the PIAA diffraction by the pupil edges is carried out using S-Huygens, with the propagation of small but arbitrary wavefront errors evaluated using SR-Fresnel.

We rely on this analytical proof to proceed with the Fresnel approximation for all our ACAD simulations. More generally, high contrast instruments are becoming more and more complex and most of them will include two mirrors (active or not) to correct for the effects of complex apertures or high amplitude aberrations or for lossless apodization. This study defines a framework for the use of the Fresnel approximation of the Huygens propagation for small deformation of the DMs. We also defined a numerical limit to separate the case where you can use this approximation or a more complex one (SR-Fresnel, S-Huygens).

\section{Numerical simulation for 1D pupil} 
\label{sect:prop_1d}

In the previous section, we analytically demonstrated that (in 1 dimension at least), the Fresnel approximation is a better approximation than the SR-Fresnel approximation for high $Z/D$ and small apodizations ($A \sim 1$). Previous works (Vanderbei et al. 2005\cite{Vanderbei_Traub05}, Pueyo et al. 2011\cite{Pueyo11}, Carlotti et al. 2011\cite{Carlotti11}{}) have studied the simulation of PIAA (high apodization, $A \rightarrow  0$ or $A \rightarrow  \infty$) for various $Z/D$ case and conclude that Fresnel is not adapted and that SR-Fresnel is a good solution for high $Z/D$. In the context of this article (assessing simulation tools for two DM propagation, including ACAD), we give only one example of the effect described in the last section, where we take a high $Z/D$: D = 2.4cm, Z = 1m and a small apodization ($|max(A^2 - 1)| = 0.1$) and compared the SR-Fresnel approximation and the Fresnel approximation to the complete Huygens integral.

\begin{figure}
   \begin{center}
\includegraphics[width=\textwidth]{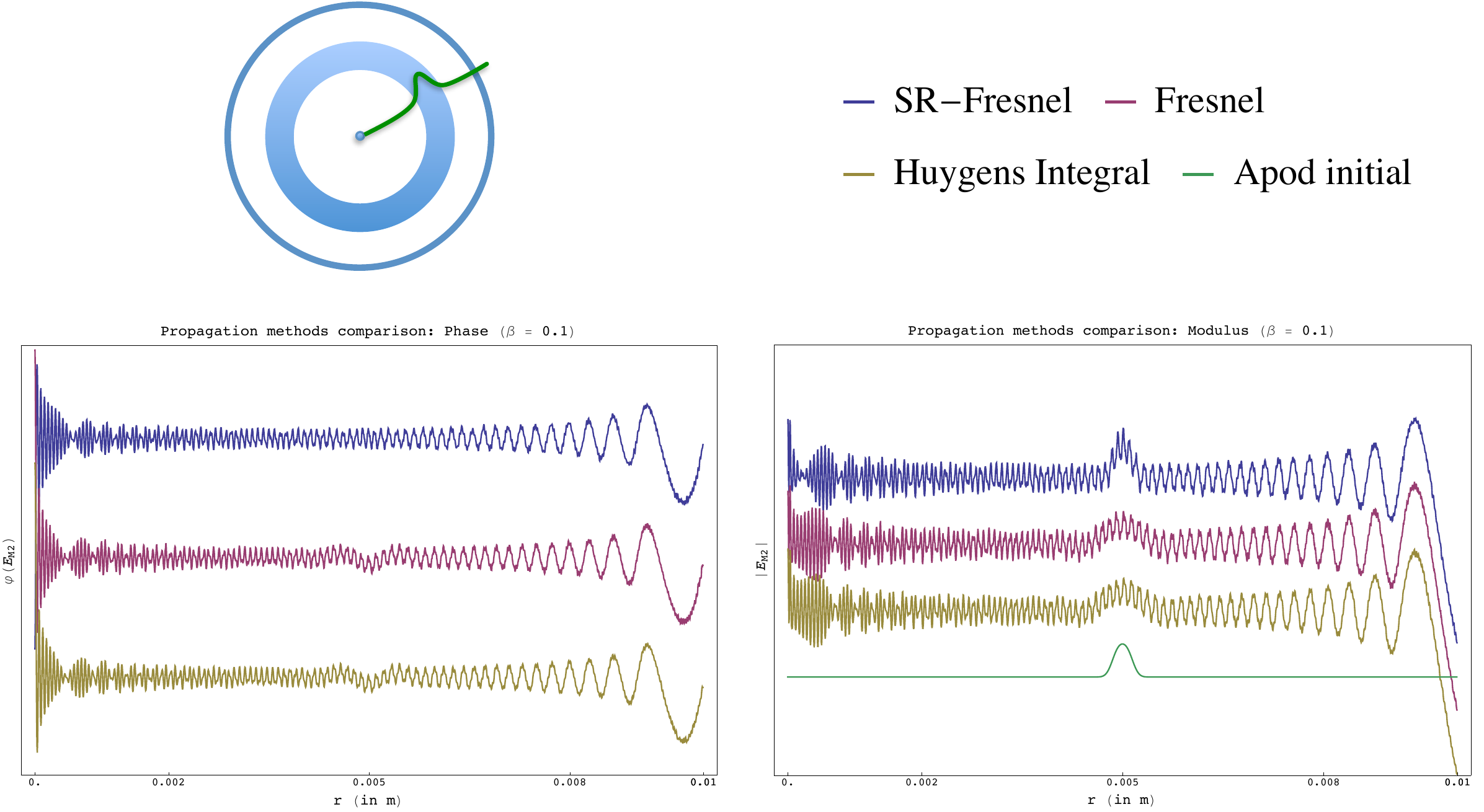}  
   \end{center}
   \caption[Numerical comparison of propagation for a given shape of the first DM]
   { \label{fig:compar_phase_ampl}
Comparison of propagation algorithms for a given axisymmetric shape of the first DM, represented in 1D by the green solid line on the top schematic and the bottom plot. The schematic at the top represents the actual shape of the 2D pupil apertures, of which the green solid line is a 1D cross section. The bottom left plot represents the phase of this field. The bottom right plot represents the modulus of the complex field $E_{M2}$. In both cases we introduced an offset in phase to separate the results of the different algorithms. } 
   \end{figure}

For computing efficiency, we decided to simulate a 1D Huygens integral, as described in Belikov et al. (2006)\cite{Belikov06}{}. In that case we can use numerical methods to calculate the ground truth to which Fresnel and SR-Fresnel will be compared. The desired apodization is presented on the schematic at the top of Figure~\ref{fig:compar_phase_ampl} in 2D. The apodization is chosen with a symmetry of revolution to be simulated in 1D. Therefore, the function we actually use in the numerical simulation is the 1D cross section describing the half pupil (represented on the same schematic green solid line).
Apart from this bump, the rest of the apodization is uniform across the pupil and $\sim 1$. We compared the results for various positions (the position in percentage of the center of the ring compared to the total radius of the pupil) and height (the parameter $\beta$ allows us to set this height to $1+\beta$, above or under the rest of the apodization) of this "bump ring".

In Figure~\ref{fig:compar_phase_ampl}, we present the simulated phase (bottom left plot) and modulus (bottom right plot) of the complex field at a distance Z of the first DM for a 1D shape represented in the green solid line. In this case, we took the ring in the center of the pupil (position = 50$\%$ of the pupil radius) and of height parameter $\beta = +0.1$ compared to the rest of the pupil. The reference Huygens integral simulated field is represented in yellow and the two approximations are plotted in blue (SR-Fresnel) and red (Fresnel). In both cases, an offset was added to separate the curves. A quick view of these plots shows that in the case of small apodizations on the first DM, the Fresnel approximation is actually closer to the Huygens integral propagation than the SR-Fresnel approximation for phase as well as for the modulus of the final complex field.

\section{Active compensation of aperture discontinuities applied to a WFIRST-AFTA-like pupil} 
\label{sect:ACAD}

In this last section, we present preliminary results of the ACAD algorithms for an WFIRST-AFTA-like pupil, represented in Figure~\ref{fig:results_ACAD} (top left). In a previous communication, Pueyo et al. (2014)\cite{Pueyo14} showed results for different aperture geometries of telescopes in space (HST, JWST) or on the ground (TMT), however without the formal justification for the Fresnel regime as described above. Using this new framework, we show how to apply the general method described in Pueyo and Norman to the WFIRST-AFTA pupil. In this paper, we place ourselves in the context of an WFIRST-AFTA-like simulation with the High-contrast imager for Complex Aperture Telescopes (HiCAT) (N'Diaye et al., 2014\cite{NDiaye14}{}). Located at the Space Telescope Science Institute (STScI), this testbed features a telescope simulator, two facesheet deformable mirrors, and a starlight suppression system, allowing us to test and validate the solutions described in this paper for the first time experimentally. Our goal is to show the applicability of our ACAD-based solutions for an WFIRST-AFTA-like pupil on HiCAT.

\subsection{Condition of our numerical simulation} 
\label{sect:numerical}

\begin{figure}
   \begin{center}
   \includegraphics[width=0.8\textwidth]{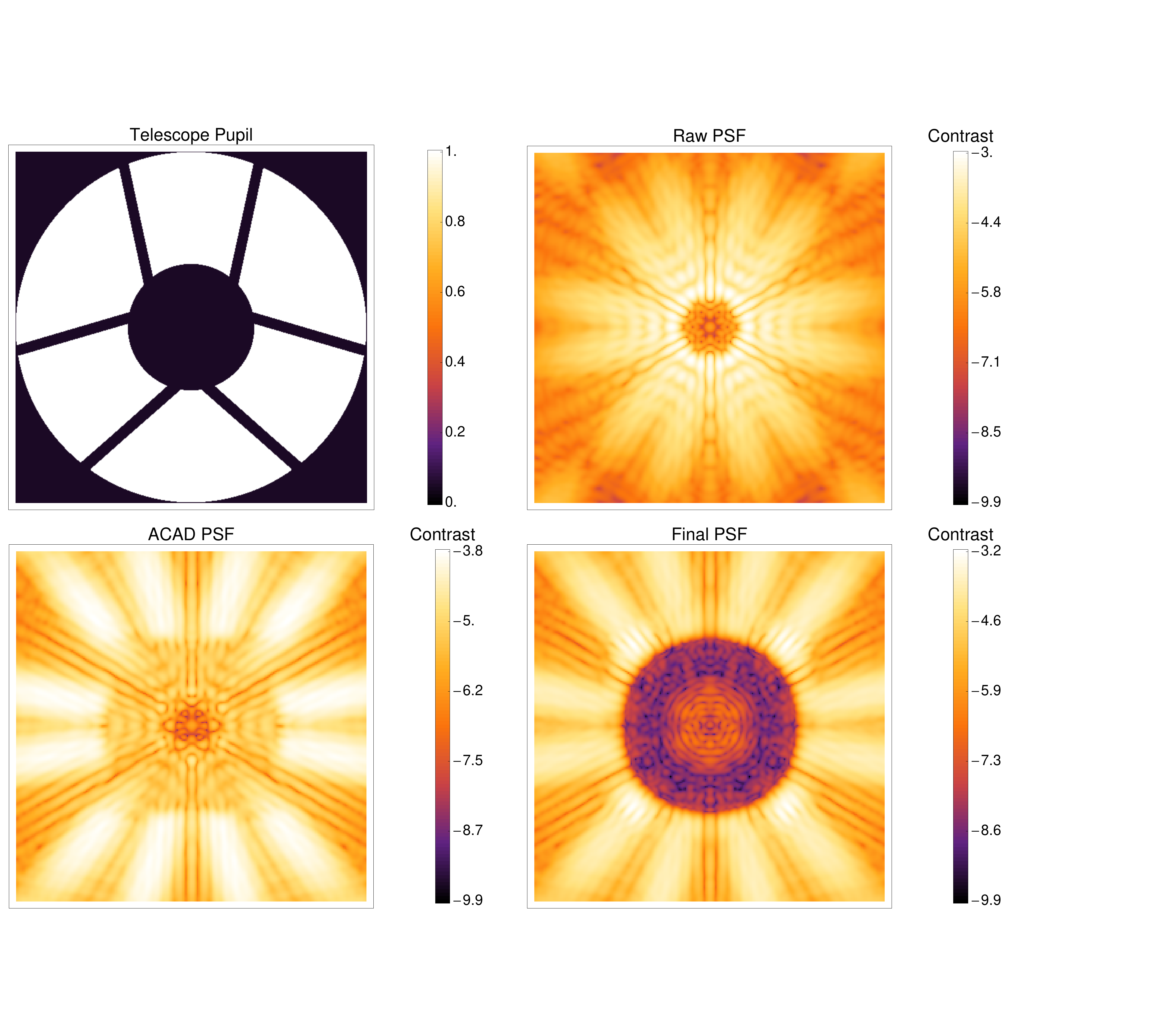}  
   \end{center}
   \caption[Focal plane image of the ACAD simulation with the WFIRST-AFTA pupil at different steps of the correction] 
   { \label{fig:results_ACAD}
Focal plane image of the ACAD simulation with the WFIRST-AFTA pupil at different steps of the correction. Top, left: The WFIRST-AFTA pupil with a 36\% central obscuration. Top, right: focal plane image obtained with this pupil with an APLC designed for a circular pupil with 36\% central obscuration and no struts, before ACAD application. Bottom, left: focal plane image obtained after application of the ACAD shape on the DMs. Bottom, right: focal plane image after ACAD and stroke minimization adjustment to find the local minimum around this shape. A dark hole is produced with a contrast better than $2.10^{-9}$.} 
   \end{figure}

We describe the parameters of the pupil, coronagraph, and deformable mirrors that are used in our numerical simulations. Our WFIRST-AFTA-like pupil is defined using a circular aperture with 36\% central obstruction (expressed in pupil radius ratio) and 4\% secondary mirror support structures, see Figure \ref{fig:results_ACAD} top left. 

We use an Apodized Pupil Lyot Coronagraph (APLC, Aime et al. 2002\cite{Aime02}{}, Soummer et al. 2005\cite{Soummer05}{}, 2011\cite{Soummer11}{}) to perform starlight diffraction suppression for a circular pupil with 36\% central obstruction and no struts. The coronagraph is here designed to produce a $10^{-9}$ raw contrast PSF dark zone ranging between 5 and 40 $\lambda$/D (inner and outer working angles [IWA, OWA]) over 10\% bandwidth, using a focal plane mask with 5 $\lambda$/D radius and a Lyot stop with 50\% central obstruction (N'Diaye et al., 2015 \cite{NDiaye15}{}). 

The parameters of the two deformable mirrors are selected to match the configuration of HiCAT on which our algorithms will first be tested. The two Boston Micromachines DMs present N=34 actuators across a D=1cm size aperture. For these two devices, we measured a maximum achievable mechanical stroke of at least 520 nm from a flat position. The DMs are separated by a distance of Z = 30 cm. This corresponds to a $Z/D$ ratio of 30 (see Section~\ref{sect:Analytical}).

\subsection{A three step correction} 
\label{sect:threestepscorrection}

The solution for the DMs is obtained in three successive steps: calculation of the ACAD solution, light propagation through the coronagraph, and DM shape adjustment using the wavefront control algorithm.

We first calculate the best possible remapping solution, hereafter named the ACAD solution, by numerically solving the Monge-Ampere equation (Equation~\ref{eq:Monge_ampereh1} or Equation~\ref{eq:Monge_ampereh2}). The details of this solver are described in Pueyo \& Norman (2013)\cite{Pueyo_Normann13}{} and Mazoyer et al.\cite{Mazoyer15}{}. To reduce the stroke applied on the actuators, we taper the discontinuities that we want to correct in the pupil by a Gaussian filter. We also multiply $A$ by a binary mask to ensure that our algorithm  will concentrate its efforts on the support structures of the secondary and to prevent it from correcting the diffraction effects due to the central obstruction or the pupil edge (leaving such operations to the coronagraph and its apodization). This is an ``open loop'' solution and our solver does not use any post-coronagraph image plane based metric.\\

In a second step, we propagate the solution of the two DM surfaces through the APLC coronagraph using the Fresnel approximation to obtain the contrast in the final image plane. We proved in the previous sections that this propagation is more accurate than the SR-Fresnel propagation.\\

Finally, we use a close loop quasi-linear algorithm called Stroke minimization (Pueyo et al., 2009 \cite{Pueyo09}{}) to adjust the DM surfaces and obtain the final contrast. Starting from the ACAD solution, this algorithm seeks a two-DM solution that improves the contrast in the image plane by 5\% at each iteration, while minimizing the stroke of the actuators. This algorithm is particularly useful in this specific case, as important strokes have already been put on the DM actuators with the ACAD solution to correct for the support structures of the secondary. In this third step, we assume a perfect estimation of the complex electric field in the focal plane for both the monochromatic and 10\% broadband light cases. Several methods have been developed to retrieve an estimate of this field\cite{Borde06,Giveon07,Baudoz06,Mazoyer14}{}, which are not specific to the 2 DMs correction. 

Due to the limited number of actuators (34 across the pupil diameter), we are only allowed to correct for a small part of the focal plane, called the dark hole (DH), of 17 $\lambda/D$ in radius. In addition, we limit the correction to an annulus of dimensions ranging between 5 $\lambda/D$ (coronagraph IWA, we have no interest in correcting the inner region) and 15 $\lambda/D$ (OWA). By somewhat reducing the DH dimensions, we ensure that the stroke minimization algorithm will not attempt to correct the strong aberrations outside the DH, created by the support structures of the secondary and not corrected by the ACAD algorithm. 

The number of iterations is variable, as the algorithm stops when the correction starts to diverge. In the current implementation of our algorithms, we apply the same weight to every actuator in this correction. Because some of these actuators are hidden behind the structures of the secondary, this might not be the optimal solution.

\subsection{ACAD on an WFIRST-AFTA like pupil: results and predictions for HiCAT} 
\label{sect:results_AFTA}
   \begin{figure}
   \begin{center}
   \includegraphics[width=0.7\textwidth]{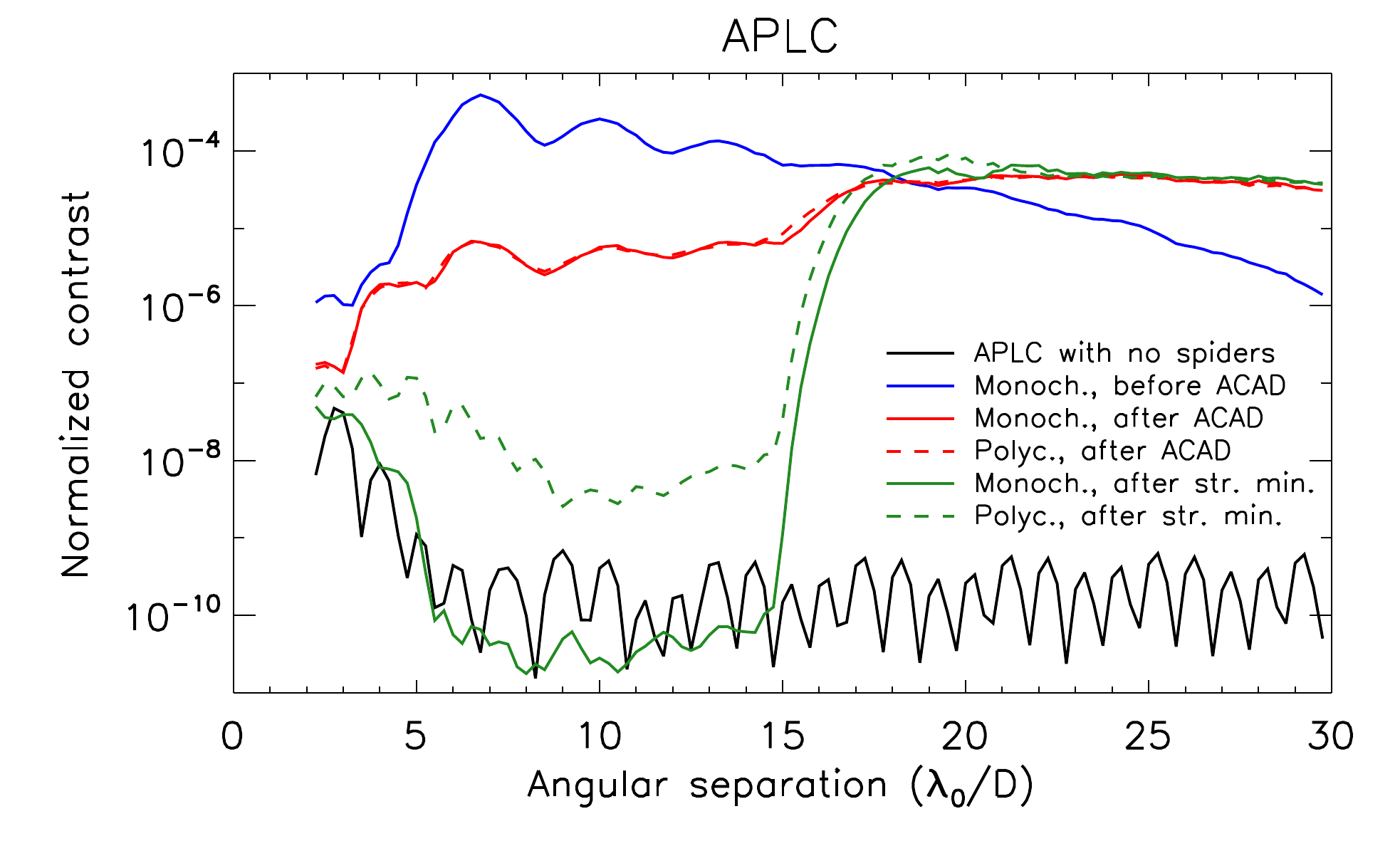}  
   \end{center}
   \caption[Radial contrast profile of the images at different stages of the ACAD + stroke minimization process for an WFIRST-AFTA like pupil.] 
   { \label{fig:results_ACAD_contrast}
Radial contrast profile of the images at different stages of the ACAD + stroke minimization process for an WFIRST-AFTA like pupil. These contrasts are normalized by the intensity peak of the PSF without coronagraph. Results in monochromatic and 10\% broadband light are represented in solid and dashed lines. The black curve represents the contrast obtained with an APLC in the case of a 36\% central obscuration pupil and no struts. The blue line shows the contrast after the introduction of the WFIRST-AFTA support structures of the secondary in the pupil. In green, the contrast in the dark hole after the ACAD solution. Finally, we present the final contrast obtained after ACAD and stroke minimization correction (red curve). A contrast level better than $6.10^{-11}$ is obtained in monochromatic light and $1.10^{-8}$ in broadband light on a 5-15 λ/D dark hole.} 
   \end{figure} 
The results of these algorithms in the image plane are presented in Figure~\ref{fig:results_ACAD}. The first panel (top left) presents the WFIRST-AFTA-like pupil that we used in this numerical simulation. The top right panel shows the effect of the structures of the secondary on the coronagraphic image produced by our APLC design. Figure~\ref{fig:results_ACAD_contrast} shows the radial profile of the contrast at different stages of the ACAD + stroke minimization process. We can see the $10^{7}$ difference in performance between the case without (black solid curve) and with (blue solid curve) these secondary structures for the same APLC. We then apply the ACAD algorithm to find the 2 DM solutions to correct for these structures. The result is presented in the bottom left panel of Figure~\ref{fig:results_ACAD}. We plot the radial profile of the contrast obtained at this stage with a solid green line in Figure~\ref{fig:results_ACAD_contrast}. Finally, we apply the Stroke minimization algorithm. Results in the focal plane are shown in the bottom right panel of Figure~\ref{fig:results_ACAD} and the performance in contrast is plotted with a solid red line in Figure~\ref{fig:results_ACAD_contrast}. We manage to generate a DH with a contrast better than $6.10^{-11}$ (black dashed horizontal line) between 5 and 15 $\lambda/D$.

The effect of polychromatic light on the Fresnel propagation have been well studied in Pueyo \& Kasdin (2007)\cite{Pueyo_Kasdin07}{}. In Figure~\ref{fig:results_ACAD_contrast}, we plot the results obtained in broadband light after the ACAD correction (green dashed line) and after the stroke minimization correction (red dashed line). We used three wavelengths covering a 10\% bandwidth. 

As expected, the performance of ACAD correction (which aims at correcting the pupil struts, independently from the incident light chromaticity) is barely altered. However, the stroke minimization correction using a contrast estimation in the image plane is more impacted by the change to polychromatic light. The contrast still remains better than $1.10^{-8}$ between 5 and 15 $\lambda/D$ over 10\% bandwidth. 

This algorithm introduces small deformations on the DMs, allowing us to keep most of the energy of an off-axis source in the PSF core. This attribute guarantees a good throughput of the overall system\cite{Mazoyer15}. We will present these results in a forthcoming paper.

   \begin{figure}
   \begin{center}
   \includegraphics[width=0.7\textwidth]{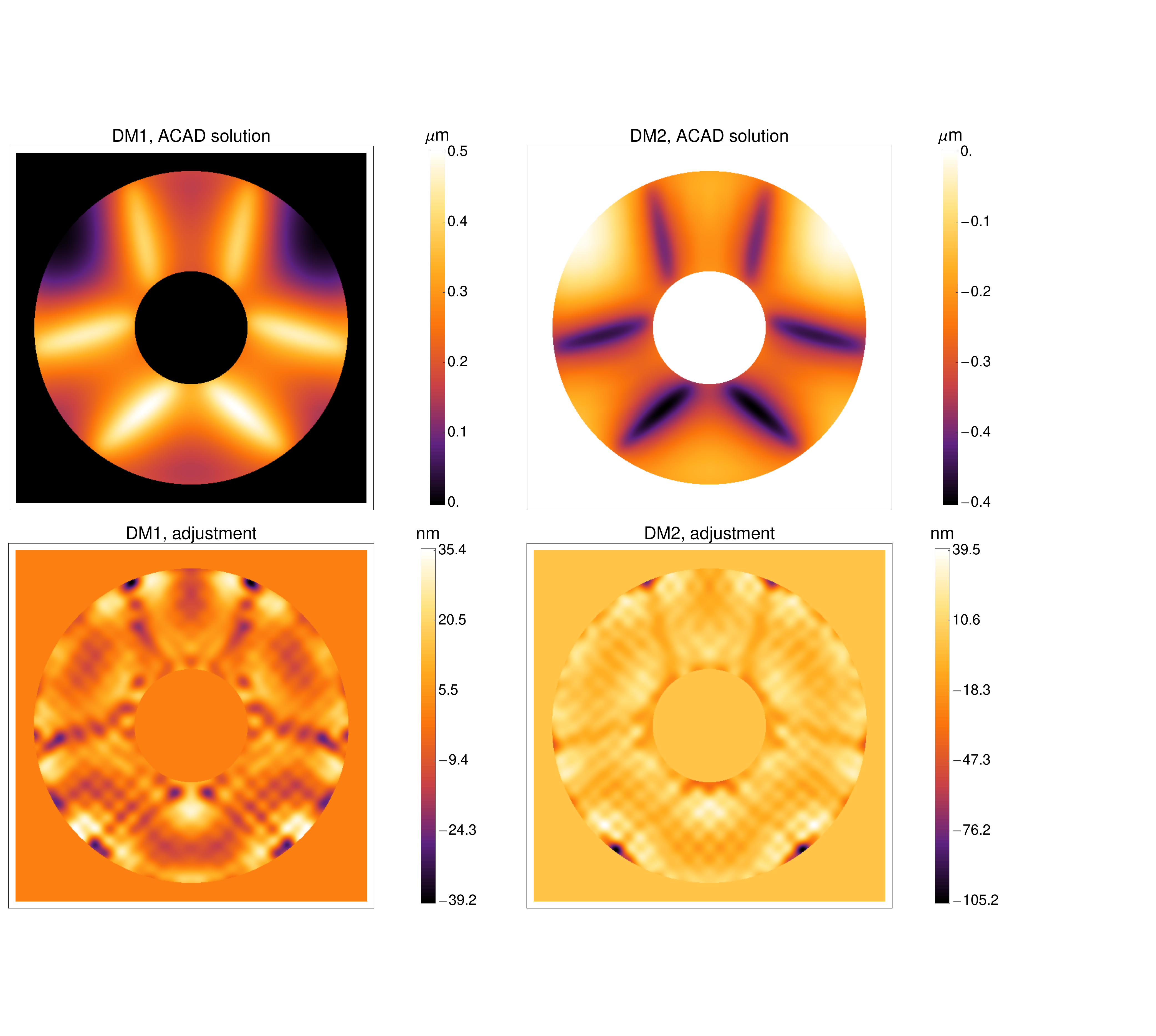} 
   \end{center}
   \caption[DM shapes at different stages of the ACAD + stroke minimization process for an WFIRST-AFTA like pupil.] 
   { \label{fig:results_ACAD_dmshape}
DM shapes at different stages of the ACAD + stroke minimization process for an WFIRST-AFTA like pupil. Top: Shapes of the ACAD solution for the DM 1 and 2. These deformations range from 0 to 380 nm. Bottom: adjustments around this shape after the stroke minimization algorithms. These deformations range from -50 to 25 nm.} 
   \end{figure} 
 
Figure~\ref{fig:results_ACAD_dmshape} shows the two DM shapes obtained after ACAD correction (top). These deformations ranged from 0 to 450 nm (mechanical) for the two DMs. The bottom part shows the adjustments to this correction introduced by the stroke minimization algorithm. The stroke involved in the second part are much more modest and ranged from -50 to 25 nm. These numbers are similar for the polychromatic case. Since we measured a maximum mechanical stroke of at least 520 nm from their flat position for the Boston micromachines DMs in HiCAT, we are confident about our ability to achieve the estimated DM strokes and produce the solutions described above on our testbed.

\section{Conclusion}
 
In the first part of this paper we proved that the numerical simulations of a two mirror propagation can be achieved more accurately using the Fresnel approximation than using the SR-Fresnel propagation in the case of a small desired apodization. We derived an analytical framework in which this approximation is true and proved that this result is not applicable for fixed mirrors apodization (PIAA), but can be used for all two DM correction techniques, where the strokes of the actuators are limited to a few $\mu$m at most.

Using this framework, the second part of this paper aimed at testing the ACAD technique with the complex pupil of WFIRST-AFTA, in the context of a future test on the HiCAT optical bench. We show that the required strokes are achievable on our DMs and that we should be able to achieve high contrast performance using the WFIRST-AFTA pupil on this testbed. 

The next steps include numerical simulation with more accurate parameters of the WFIRST-AFTA mission (larger DMs with more separation) and with other coronagraphs (e.g. vortex).

\section*{Annex A} 

Using Equation~\ref{eq:S}, we can derive:
\begin{equation}
	\label{eq:Sfresnel}
	\dfrac{S(z,y,x_2,y_2)}{Z} = \left(1 + \dfrac{1}{2}\left(\dfrac{x-x_2}{h_1-h_2}\right)^2 + \dfrac{1}{2}\left(\dfrac{y-y_2}{h_1-h_2}\right)^2\right) \dfrac{h_1-h_2}{Z} + \mathcal{O}(\left(\dfrac{D}{Z}\right)^4)
\end{equation}
Therefore 
\begin{eqnarray}
	\label{eq:Qfresnel3}
	\dfrac{Q(z,y,x_2,y_2)}{Z} &=& 2\dfrac{h_1 - h_2}{Z} + \dfrac{1}{2} \left(\left(\dfrac{x-x_2}{Z}\right)^2 + \left( \dfrac{y-y_2}{Z}\right)^2 \right) \dfrac{Z}{h_1-h_2} + \mathcal{O}(\left(\dfrac{D}{Z}\right)^4)\\
	  &=&\dfrac{h_1 - h_2}{Z} + \dfrac{1}{2} \left(\left(\dfrac{x-x_2}{Z}\right)^2 + \left( \dfrac{y-y_2}{Z}\right)^2 \right) \dfrac{1}{1 +\dfrac{(h_1-Z) - h_2}{Z}} + \mathcal{O}(\left(\dfrac{D}{Z}\right)^4) \nonumber
\end{eqnarray}
The next step to express the order of the Fresnel approximation in $\dfrac{D}{Z}$ is to calculate the order $\alpha$ of the following%
\begin{eqnarray}
 \dfrac{1}{1 +\dfrac{(h_1-Z) - h_2}{Z}} &=& 1 + \mathcal{O}(\dfrac{(h_1-Z) - h_2}{Z}) \nonumber \\
                                        &=&  1 + \mathcal{O}(\left(\dfrac{D}{Z}\right)^\alpha)
\end{eqnarray}
In this context the approximation is driven by the quantities that one is integrating over. We therefore have to evaluate ($h_1 -Z$) and $h_2$, using Equation~\ref{eq:Monge_ampereh1} and Equation~\ref{eq:Monge_ampereh2}, which constrain their second derivative. Let us define $M = |max(A^2 - 1)|$ and $m = |max(1/A^2 -1)|$. If we use Equation~\ref{eq:Monge_ampereh1} and Equation~\ref{eq:Monge_ampereh2} in 1 dimension (1D), we have $\left| \dfrac{\partial^2 h_1}{\partial x^2}\right| \leq m/2Z$ and $\left|\dfrac{\partial^2 h_2}{\partial x^2}\right| \leq M/2Z$. To integrate these inequalities, we have to ensure, for both DMs, that there is at least one point of the surface were the derivative is zero and one point where the surface is zero. Because the axis of the system has to be conserved, we have at least: $\dfrac{\partial^2 h_2}{\partial x^2}\bigg|_{(D/2,d/2)} = \dfrac{\partial^2 h_1}{\partial x^2}\bigg|_{(D/2,d/2)}= 0$. We have no piston on both DMs, so there exists at least one point on each of them where $h_1(x_{1},y_{1}) - Z = 0$ and $h_2(x_{2},y_{2}) = 0$. Using the proper integration bounds, we can finally conclude: 
\begin{eqnarray}
	\label{eq:approx_h1}
	|h_1 - Z|  &\leq& mD^2/(2Z) \nonumber\\
	\dfrac{h_1 - Z}{Z}      &=& \mathcal{O}(\dfrac{m}{2}\left(\dfrac{D}{Z}\right)^2)
\end{eqnarray}
and
\begin{eqnarray}
	\label{eq:approx_M}
	|h_2|  &\leq& MD^2/(2Z) \nonumber\\
	\dfrac{h_2}{Z}      &=& \mathcal{O}(\dfrac{M}{2}\left(\dfrac{D}{Z}\right)^2)
\end{eqnarray}
As a consequence: 
\begin{equation}
	\label{eq:Qfresnel4}
\dfrac{Q(z,y,x_2,y_2)}{Z} = 2\dfrac{h_1 - h_2}{Z} + \dfrac{1}{2} \left(\left(\dfrac{x-x_2}{Z}\right)^2 + \left( \dfrac{y-y_2}{Z}\right)^2 \right) + \mathcal{O}((\dfrac{m+M}{2})\left(\dfrac{D}{Z}\right)^2) + \mathcal{O}\left(\dfrac{D}{Z}\right)^4)
\end{equation}
when both $m \ll Z/D$ and $M \ll Z/D$, the Fresnel approximation (Equation~\ref{eq:Qfresnel2}) is actually an $\mathcal{O}(\left(\dfrac{D}{Z}\right)^4)$ approximation and is more accurate than the SR-Fresnel one. \\

\acknowledgments 
This material is based upon work carried out under subcontract \#1496556 with the Jet Propulsion Laboratory funded by NASA and administered by the California Institute of Technology. The authors also want to thank John Krist (Jet Propulsion Laboratory) for his critical help in this research. 


\bibliography{article_jatis}   
\bibliographystyle{spiejour}   


\vspace{2ex}\noindent{\bf Johan Mazoyer} is currently a postdoc at the Space Telescope Science Institute. He graduated from the \'E{}cole polytechnique (Paris, France) in 2011 and received a PhD in Astronomy and Astrophysics from Paris Diderot University/Paris Observatory (France) in 2014. His research interests lie both in the development of innovative instruments for imaging close circumstellar environments (planets or dust) and in the analysis of high contrast images. 

\vspace{2ex}\noindent\textbf{Laurent Pueyo} is currently an associate astronomer at Space Telescope Science Institute. He is developing innovating correction and post-processing techniques for high contrast coronagraphic instruments. He is also carrying out observations of circumstellar environments with several instruments (GPI, HST, P1640).

\vspace{2ex}\noindent\textbf{Colin Norman} has been a professor of physics and astronomy at Johns Hopkins University and astronomer at the Space Telescope Science Institute since 1984. He works in both theoretical and experimental astrophysics. His most recent observational project concerns the analysis and interpretation of the Chandra deep fields. He is currently principal investigator of the Hubble Origins Probe. His recent theoretical work concerns the magnetic collimation of astrophysical jets.

\vspace{2ex}\noindent{\bf Mamadou N'Diaye} is a STScI postdoc with expertise in high-contrast imaging and spectroscopy of circumstellar environments (exoplanets, disks), developing cutting-edge concepts in starlight suppression, wavefront sensing and control. Highly proficient in coronagraphy with several refereed publications, he has recently proposed a Lyot coronagraph for the observation of planets $10^{10}$ fainter than their host star with segmented telescope apertures. He is an expert in Zernike wavefront sensing approaches and kingpin on ZELDA, a Zernike sensor for the real-time measurements of coronagraphic aberrations on VLT/SPHERE. He has also performed a substantial portion of the design and assembly of HiCAT, the high-contrast testbed for arbitrary telescope apertures in Baltimore.

\vspace{2ex}\noindent{\bf Roeland van der Marel} is an astronomer at the Space Telescope Science Institute (STScI) and an adjunct professor at nearby Johns Hopkins University. At STScI he is the Mission Lead for WFIRST-AFTA. He is a frequent user of the Hubble Space Telescope, and an expert on black holes and the structure of galaxies.

\vspace{2ex}\noindent\textbf{R\'e{}mi Soummer} is an astronomer at the Space Telescope Science Institute. He develop novel coronagraph designs, speckle statistics theories, and innovative high-fidelity modeling algorithms for high-contrast imaging.  At STScI, he is currently leading the  Telescopes group in the Instruments division and the Russell B. Makidon Optics Laboratory.

\listoffigures

\end{document}